# A Simple Method for Determining Center Shift and Spectral Area of a Mössbauer Spectrum

Stanisław M. Dubiel[1*] and Jan Żukrowski[2]

[1]Faculty of Physics and Applied Computer Science, [2]Academic Center for Materials and Nanotechnology, [1,2]AGH University of Science and Technology, PL-30-059 Kraków, Poland

**Abstract**

A very simple method for determining the center (isomer) shift, *CS*, and the spectral area, *A*, of a Mössbauer spectrum is outlined. Its applicability is demonstrated in two examples viz. pyrite and a ternary sigma-phase Fe-Cr-Ni compound. Sets of the spectra recorded in the temperature interval of 78-295 K for the former, and 78-295 K for the latter were analyzed with the simple and a standard method. *CS(T)*- and *A(T)*-data obtained with both ways of the spectra analysis were analyzed in terms of the Debye model. The determined values of the Debye temperatures agree well within the error limit with each other proving that this very simple method is suitable for a quick determining of *CS* and *A* spectral parameters.

* Corresponding author: Stanislaw.Dubiel@fis.agh.edu.pl

## 1. Introduction

Two spectral parameters characteristic of a Mössbauer spectrum, (1) center shift, *CS*, also known as an isomer or chemical shift, and (2) spectral area, *A*, are of interest as far as lattice vibrations are concerned. Concerning (1), its temperature dependence, *CS(T)*, permits determination of the Debye temperature, $T_D$, via the following formula:

$$CS(T) = IS(0) + SOD(T) \qquad (1)$$

Where *IS* stands for the isomer shift and *SOD* is the so-called second order Doppler shift i.e. a quantity related to a non-zero mean value of the square velocity of vibrations, $<v^2>$, hence a kinetic energy of vibrating atoms. In the approximation of the Debye model, and assuming that *IS* hardly depends on temperature, so it can be ignored [1], the *T*-dependence of *CS* practically depends on the second term in Eq. (1) which is related to $T_D$ via the following equation [2]:

$$CS(T) = IS(0) - \frac{3k_B T}{2mc}\left(\frac{3T_D}{8T} + 3\left(\frac{T}{T_D}\right)^3 \int_0^{T_D/T} \frac{x^3}{e^x - 1} dx\right) \qquad (2)$$

Here *m* is the mass of the Fe atom, $k_B$ is the Boltzmann constant, *c* is the speed of light, and $x = \frac{\hbar\omega}{2\pi k_B T}$ ($\omega$ being frequency of vibrations).

Spectral area, *A*, in turn, is of importance as it is proportional to the recoil-free fraction, *f*, which is related to the average value of the square amplitude of vibrations, $<x^2>$, hence a potential energy of vibrating atoms (assuming the harmonic vibrations) by the following expression:

$$f = \exp[-(\frac{E_\gamma}{\hbar c})^2 \langle x^2 \rangle] \qquad (3)$$

Where $E_\gamma$ stands for the energy of $\gamma$-rays (e. g. 14.4 keV for $^{57}$Fe and 23.8 keV for $^{119}$Sn) and *c* is the light velocity.

Assuming the lattice vibrations can be described by the Debye model, the recoil-free fraction can be related to the Debye temperature by the following formula:

$$f = \exp[-\frac{6E_R}{k_B T_D}\{\frac{1}{4}+(\frac{T}{T_D})^2 \int_0^{T_D/T} \frac{x}{e^x - 1} dx\}]  \qquad (4)$$

$E_R$ is here for the recoil-free energy, $E_R = \frac{E_\gamma^2}{2mc^2}$.

Determining of *CS* and/or *A* from a spectrum is an easy task if the spectrum is simple i.e. it is either in the form of a single line, a doublet or a sextet. It is also a rather simple task if the spectrum is well-resolved, the number of lattice sites occupied by probe nuclei and their population on these sites are known. However, if the investigated sample has a complex crystallographic structure i.e. with several sub lattices or lower-than-cubic symmetry and, in addition, but not necessarily, it is weakly magnetic, determination of the correct values of these parameters may be a challenging issue. Good examples of such kind of spectra are those measured on so-called Frank-Kasper phases, also known as topologically close-packed (TCP) phases e. g. σ, λ, ζ, η and other [3]. The best known case of them is sigma (σ). It has a tetragonal unit cell with 30 atoms distributed over five different lattice sites [4]. In addition, at low temperatures, it may be weakly magnetic e. g. Fe-X (X=Cr, V, Re, Mo) just to name examples of a binary Fe-based cases [5-8]. Consequently, in order to properly analyze a spectrum measured on σ a full Hamiltonian must be included into the fitting procedure.

The simplest possible method of determining *CS* and *A* is outlined in this article and applied in two examples. The advantages of the new method are as follows: (1) prior knowledge of the number and types of components (singlet, doublet, sextet) or their relative contribution is not needed, (2) it is extremely quick i.e. much faster than the traditional way of the spectra analysis because it does not involve an iteration procedure.

## 2. The simple method

### 2.1. Center shift

The center shift of a Mössbauer spectrum, *CS*, is calculated as an average velocity weighted over the number of counts in each channel i.e. using the following formula:

$$CS = \frac{\sum_{k=1}^{n}(<b> - Y^k)V^k}{\sum_{k=1}^{n}(<b> - Y^k)} \qquad (5)$$

Where:

$<b> = \frac{\sum_{k=1}^{n1} Y^k + \sum_{k=n-n1+1}^{n} Y^k}{2n1}$ is the background of the spectrum, $Y^k$ stands for the number of counts in the *k*-th channel, $V^k$ is the source velocity of the *k*-th channel, *n* is the number of channels per spectrum and *n1* indicates the number of channels taken for the calculation of the <b> parameter (the first *n1* and the last *n1* channels out of *n* are used to calculate <b>).

Errors of <b>, Δ<b>, and of *CS*, ΔCS, are calculated based on the following formulas, respectively:

$$\Delta<b> = \left(\frac{\sum_{k=1}^{n1}(Y^k - <b>)^2 + \sum_{k=n-n1+1}^{n}(Y^k - <b>)^2}{2n1(2n1-1)}\right)^{1/2} \qquad (6)$$

$$\Delta CS = CS \left(\left|\frac{\sum_{k=1}^{n}(\Delta<b> + (<b> - Y^k)^{\frac{1}{2}}))V^k}{\sum_{k=1}^{n}(<b> - Y^k)V^k}\right| + \left|\frac{\sum_{k=1}^{n}(\Delta<b> + (<b> - Y^k)^{\frac{1}{2}}))}{\sum_{k=1}^{n}(<b> - Y^k)}\right|\right) \qquad (7)$$

### 2.2 Spectral area

As the value of the spectral area (area under the spectrum) depends on the number of counts (statistics), to take this fact into account, a normalized spectral area, *A = A' / <b>*, is calculated where:

$$A' = A2' + (<b> - Y^n) | (V^n - V^{n-1}) | /2 \tag{8a}$$

$$A2' = A1' + \sum_{k=2}^{n-1} (<b> - Y^k) | (V^{k+1} - (V^{k-1}) | /2 \tag{8b}$$

$$A1' = (<b> - Y^1) | (V^2 - V^1) | /2 \tag{8c}$$

An error of the normalized spectral area, $\Delta A'$, is calculated using as follows:

$$\Delta A = A' \left( |\frac{\Delta A'}{A'}| + |\frac{\Delta <b>}{<b>)}| \right) \tag{9a}$$

$$\Delta A' \cong \sum_{k=1}^n (\Delta <b> + (Y^k)^{1/2}) | (V^{k+1} - (V^{k-1}) | /2 + \sum_{k=1}^n (<b> - Y^k) \Delta V$$

$$\Delta V \cong 0.0005 * V^{max}/n \tag{9b}$$

The above equations were encoded using the FORTRAN programming language. The code has been complied for the platform of the current Windows version.

## 3. Examples

Two cases demonstrating the application of the method are presented. One depicts the pyrite, $FeS_2$, which can be regarded as an "easy case", and another one has to do with a sigma (σ) FeCrNi intermetallic alloy, a "difficult case" as far as the standard spectrum analysis is concerned.

### 3.1. Pyrite

Unit cell of pyrite, $FeS_2$, has a cubic symmetry but Fe atoms have a slightly distorted octahedral symmetry [9], hence they experience an electric field gradient. Consequently, two hyperfine spectral parameters viz. *CS* and a quadrupole splitting, *QS,* are needed to correctly analyze its Mössbauer spectrum. Room temperature values of *CS*, relative to that of metallic iron, spread between 0.310 mm/s and 0.329 mm/s and those of *QS* between 0.603 and 0.624 mm/s [10, 11]. The spread can be explained in terms of variations in compositions (different impurities and their content) and conditions of formation. In addition, differences in Mössbauer set-ups and spectra analysis procedures can contribute to the spread. In this study [57]Fe-site spectra were measured on a powdered sample of a natural $FeS_2$ in a transmission mode. 14.4 keV gamma rays were supplied by a Co/Rh source. Examples of spectra recorded at different temperatures are shown in Fig. 1. The doublet-like structure exists down to 4.2 K, but a line broadening is observed at low temperatures related to a weak magnetism of $FeS_2$ [9].

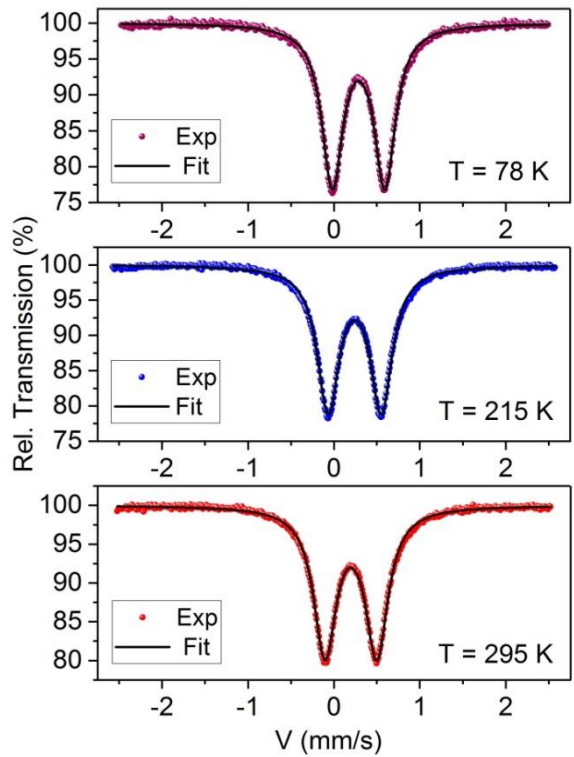

Fig. 1 Mössbauer spectra recorded on $FeS_2$ at various temperatures. Solid lines stand for the best fit of a doublet to the data.

### 3.1.1 Determining *CS*

#### 3.1.1.1 Standard approach

The spectra were fitted to a doublet with *CS*, *QS* and the linewidth, *G*, treated as free parameters. Their values at 295 K were 0.305(1) mm/s (relative to α-Fe), 0.604(1) mm/s and 0.26(1) mm/s, respectively. The temperature dependence of *CS* obtained with this fitting procedure is illustrated in Fig. 2a.

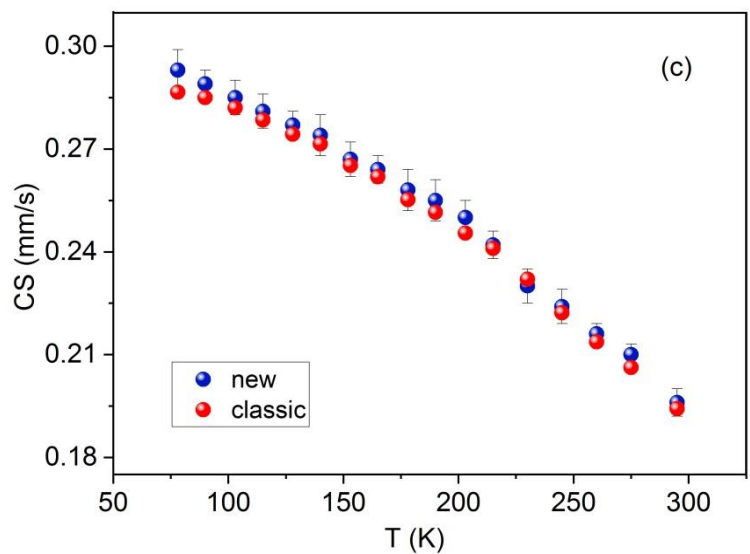

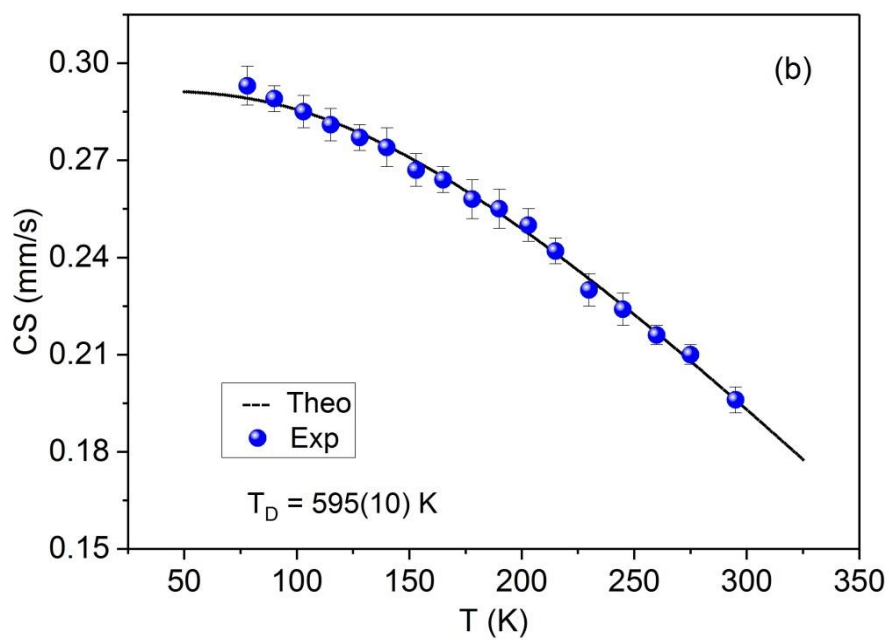

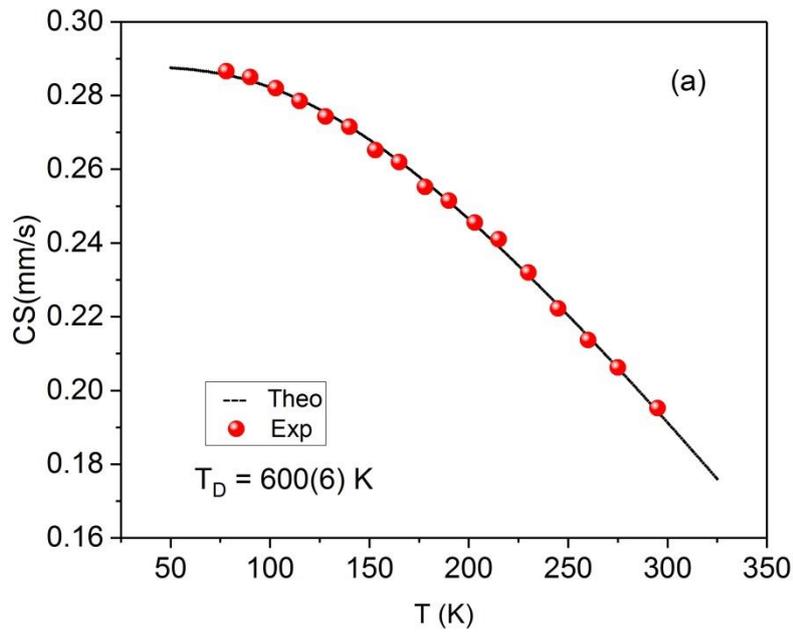

Fig. 2 Temperature dependences of the center shift, CS, as obtained with: (a) standard approach, (b) new approach and (c) comparison of both. The best-fit curve to the data in terms of eq. (2) is marked by solid lines. Note: CS-values are given relative to a Co/Rh source.

The CS(T) data were fitted to eq.(2), yielding the value of 600(6) K for the Debye temperature, $T_D$.

### 3.1.1.2. Simple approach

The spectra analysis in terms of the simple method yielded the CS-data shown in Fig. 2b. The solid line represents the best fit of this data to eq. (2). The obtained value of $T_D$ = 590(10) K, agrees within the error limit with the one found with the standard approach. The comparison of the CS(T) data obtained with both methods is shown in Fig. 2c. A good match between the two sets of the data exists for all temperatures.

This proves that the new approach to the analysis of the spectra gives the correct values of *CS*.

It is of interest to compare the $T_D$-values obtained in this study with those obtained previously by analyzing the Mössbauer spectra. Thus, Nishihara and Ogawa reported 610(15) K obtained from the analysis of *CS(T)* measured between 77 K and 292 K [12], Kansy et. al [13] found 568(20) K for a natural sample of $FeS_2$ based on the *CS(T)*-data measured in the *T*-range between 290 K and 430 K, Polyakov et al. [11] measured a synthesized sample between 90 K and 295 K and using the *CS(T)*-data arrived at the value of 551(8.5) K. Finally, the value of $T_D$=636(5) K was determined for a natural sample by analyzing the temperature dependence of the *f*-fraction [14].

### 3.1.2 Determining spectral area

### 3.1.2.1 Standard approach

Effective thickness, $\tau$, a parameter proportional to the spectral area, *A*, was determined by fitting the spectra to a doublet using the integral transmission method. In practice, one uses a normalized effective thickness, $\tau/\tau_o \sim A/A_o$, where $\tau_o$ and $A_o$ are effective thickness and the spectral area, respectively, found for the spectrum measured at the lowest temperature (here 78 K). Temperature dependence of $A/A_o$ obtained with this fitting procedure from the spectra recorded on $FeS_2$ is illustrated in Fig. 3. This data can be used to determine, via Eq. (4), the Debye temperature, $T_D$, which in this case is equal to 274(4) K.

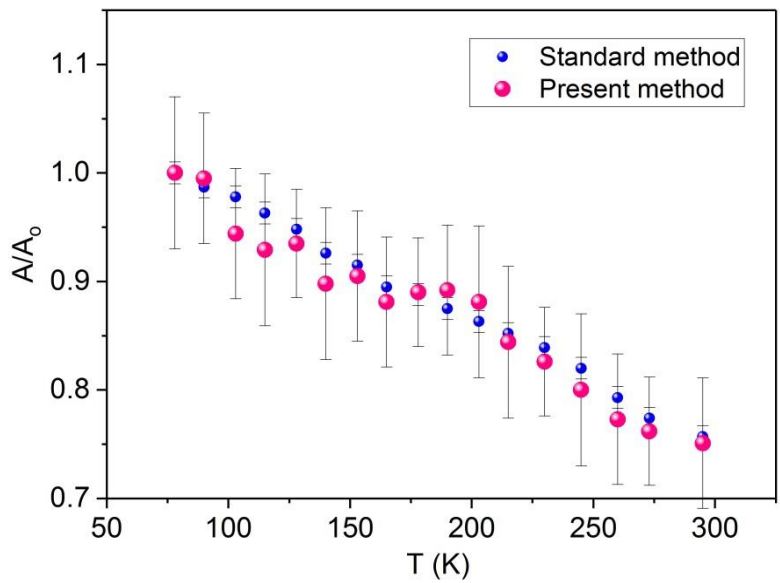

Fig. 3 Temperature dependence of the normalized spectral area, $A/A_o$, as found with the standard (common) and the new methods. Small error bars are for the standard method.

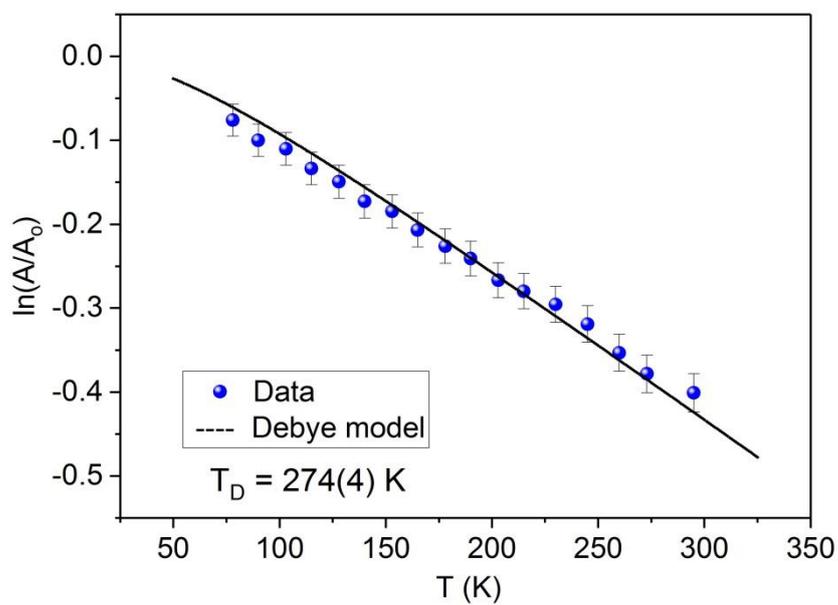

Fig. 4 Dependence of $ln(A/A_o)$ on temperature, $T$, as found analyzing the $FeS_2$ spectra with the standard method. The straight line represents the best fit to the data in terms of Eq.(4).

### 3.1.2.1 New approach

Determining the background of the spectrum, $<b>$, is important in this method as both $CS$ and $A$ values are $<b>$-sensitive, especially when the statistical quality of the measured spectrum is poor and/or the maximum velocity is not high enough. In our case – see Fig. 1, the statistical quality of the spectra is high, and the maximum velocity was correctly chosen. Consequently, as shown in Fig. 5, the $A$-values obtained for $n1=15$ are the same as those for $n1=2$.

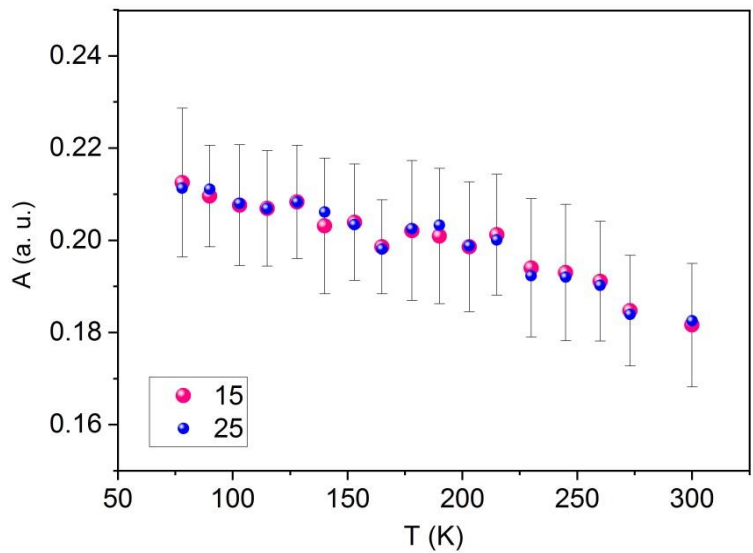

Fig. 5 Temperature dependence of the spectral area, $A$, as found from the analysis of the $FeS_2$ spectra using the new method for $n1=15$ (big symbols) and $n2=25$ (small symbols).

The *A*-values obtained with the simple method of spectra analysis are in line with those found with the standard method – see Fig. 3. Consequently, as can be seen in Fig. 6, the value of the Debye temperature retrieved by fitting this data to Eq.(4) agrees within the error limit with that found using the standard method.

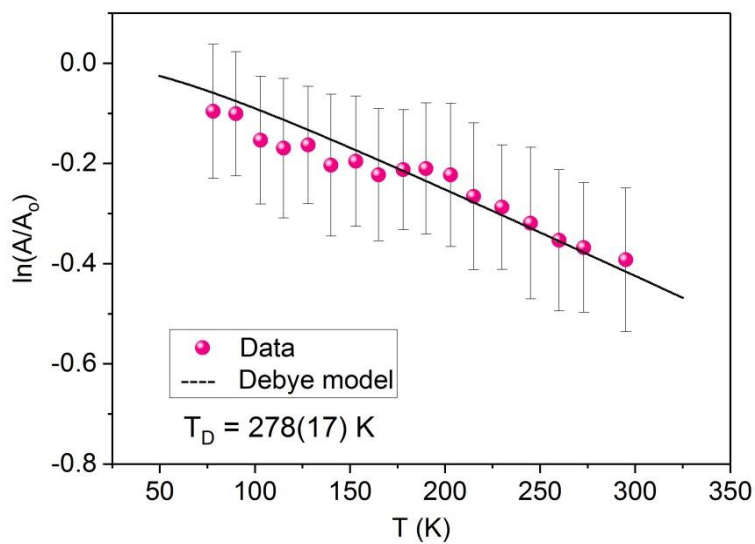

Fig. 6  *T*-Dependence of *ln(A/A$_o$)* as found analyzing the FeS$_2$ spectra with the simple method. The straight line represents the best fit to the data in terms of Eq.(4).

### 3.2. Sigma phase

$^{57}$Fe-site spectra were recorded in a transmission mode at different temperatures (78-295 K) on a σ-phase Fe$_{0.525}$Cr$_{0.455}$Ni$_{0.020}$ intermetallic compound. Examples of them are shown in Fig. 7

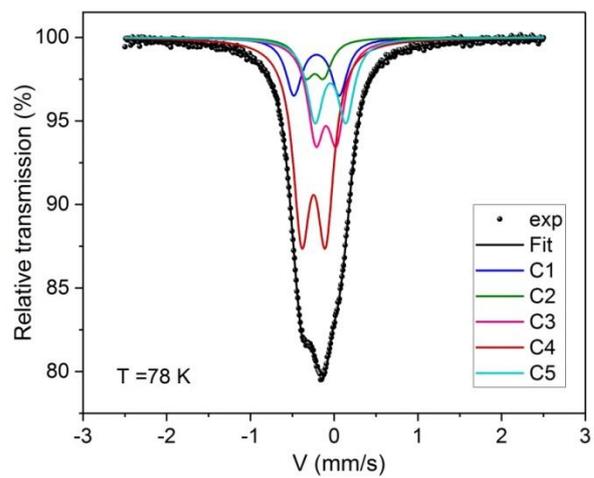
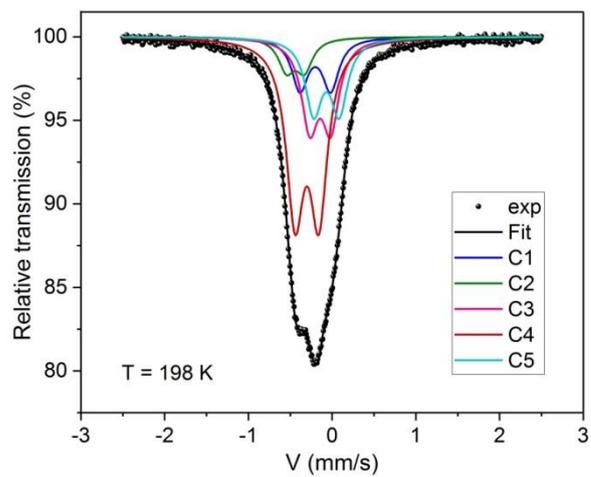
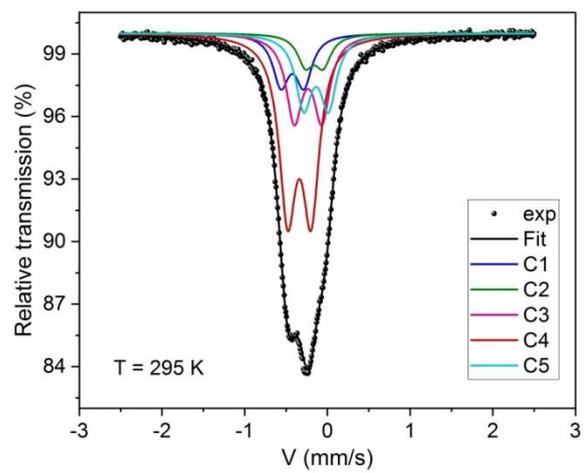

Fig. 7 Examples of the spectra recorded on the sample of σ-Fe$_{0.525}$Cr$_{0.455}$Ni$_{0.020}$ at various temperatures shown. Five sub spectra related to five lattice sites are indicated.

The magnetic ordering temperature determined with the magnetization measurements is 48 K [15], so the presented here spectra were recorded in a paramagnetic state. This means that the proper analysis of the spectra requires inclusion of the Coulomb monopole interactions and the quadrupole interactions. Furthermore, each spectrum has to be fitted into five sub spectra as Fe atoms are present on all five lattice sites what was revealed by neutron diffraction (ND) experiment. Last, but not least, a relative contribution of the sub spectra has to be known (it was determined by the ND experiment) because sigma-phase has no fixed stoichiometry, so the population of atoms on particular sites is characteristic of a given composition. In these circumstances the correct analysis of the spectra is very challenging and requires a great skill in their analysis.

### 3.2.1 Determining *CS*:

### 3.2.1.1 Standard approach

Following the above described conditions, the spectra were analyzed in terms of five components with fixed relative contributions as followed from the ND experiment. Namely, the spectra were fitted to five doublets. The average values of *CS*, *<CS>*, were determined as the weighted values i.e. *<CS>*=$\Sigma_k CS_k P_k$ where *k=1-5*, $P_k$ is the relative abundance of the *k*-th component. The obtained *<CS>(T)*-dependence can be seen in Fig. 4a. The fitting of the data to Eq. (2) yielded the value of 437(7) for the Debye temperature.

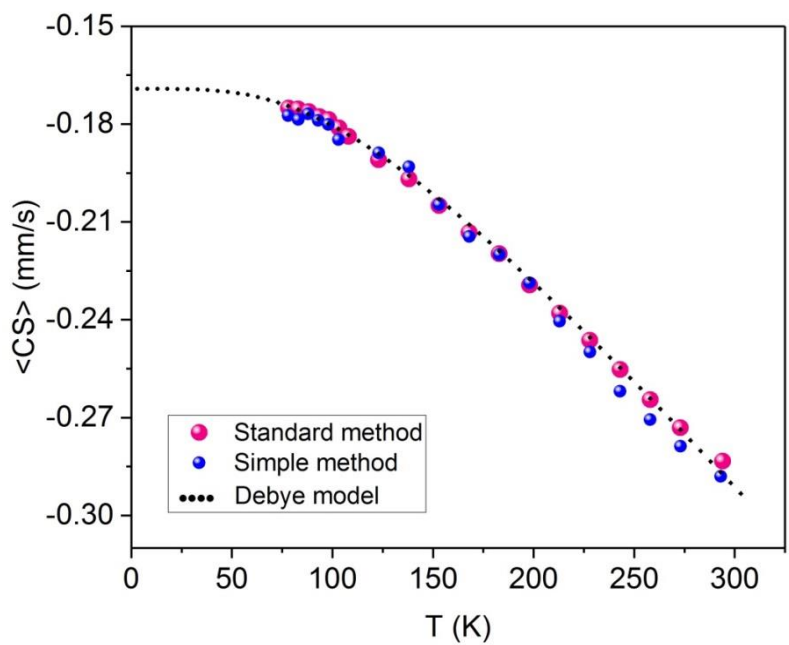

Fig. 8 Temperature dependence of the average center shift, <CS>, as found with the standard and the simple (new) method. The fit of the data to eq. (2) is marked by the dotted line. Error of <CS> determined with the new method lies in the range between ±0.003 and ±.006 mm/s while that determined with the standard method in the range of ±0.001 - ±0.002 mm/s.

#### 3.2.1.2 Simple approach

The <CS>-data determined by this method are also presented in Fig. 8. It can be clearly seen that they are very similar to those obtained with the standard method. Their analysis in terms of Eq. (2) gave the value of 450(12) K for the Debye temperature. This value agrees well within the error limit with the corresponding value found with the standard method of analysis.

### 3.2.2 Determining spectral area:

#### 3.2.2.1 Standard approach

The applied procedure of the spectra analysis was based on the transmission integral approach, and it yielded also a quantity proportional to the spectral area, $A$. The temperature dependence of its normalized value, $A/A_o$, in the semi logarithmic scale, is presented in Fig. 9. The fit of the data to Eq. (4) yielded for the Debye temperature the value of 379(5) K.

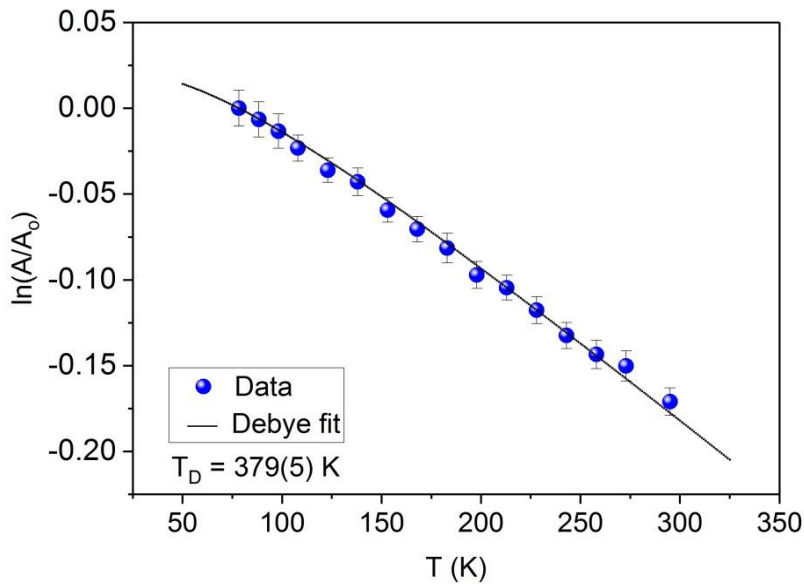

Fig. 9 $T$-Dependence of $ln(A/A_o)$ as reveled by analyzing the $\sigma$-FeCrNi spectra with the standard method. The straight line is for the best fit to the data in terms of Eq.(4). $T_D$ marks the Debye temperature.

### 3.2.2.2 Simple approach

Temperature dependence of the relative spectral area, $A/A_o$, presented in the semi logarithmic scale is illustrated in Fig. 10. The data were fitted to Eq. (4) yielding the value of 367(30) K for $T_D$. This value agrees very well with the corresponding one obtained using the standard method of the spectra analysis. Noteworthy, the values of $T_D$ obtained from the spectral area are significantly lower than the ones found from the center shift. This is known across the literature.

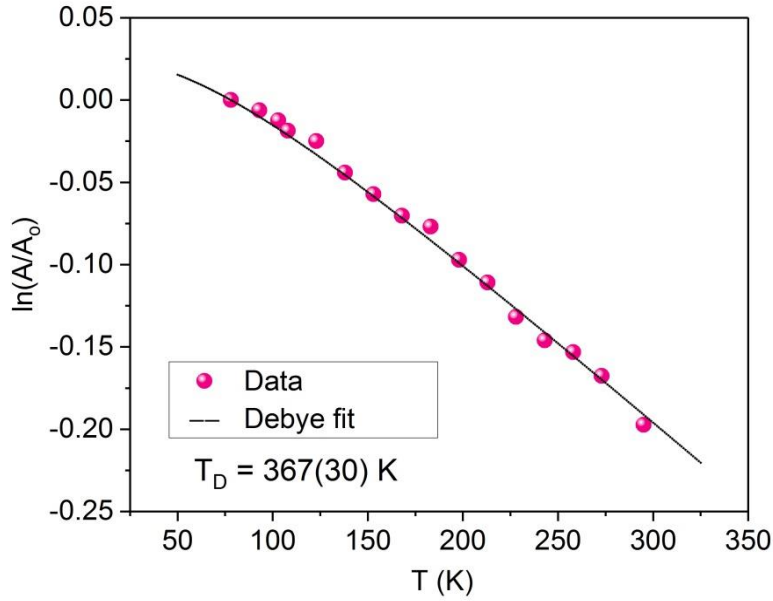

Fig. 10 *T*-Dependence of *ln(A/A_o)* as reveled by analyzing the σ-FeCrNi spectra with the simple method. The straight line is for the best fit to the data in terms of Eq.(4). $T_D$ indicates the Debye temperature.

## 4. Summary

The most simple method is outlined for determining the center (isomer) shift, *CS*, and the spectral area, *A*, of a Mössbauer spectrum. Its usefulness has been exemplified by an analysis of two sets of the spectra viz. measured in the temperature range of 78-295 K on a powder sample of a natural pyrite, $FeS_2$ and on a powder sample of a sigma-phase $Fe_{0.525}Cr_{0.455}Ni_{0.020}$ intermetallic compound. For comparison, all the recorded spectra were also analyzed using a standard procedure i.e. the spectra of $FeS_2$ were fitted to one doublet, and those of σ-$Fe_{0.525}Cr_{0.455}Ni_{0.020}$ to five doublets related to five different lattice sites. The obtained sets of the *CS(T)*- and *A(T)*-data were analyzed in terms of the relevant Debye model yielding values of the Debye temperature, $T_D$. For the pyrite $T_D$=590(10) K for the simple method and $T_D$=600(6) K for the standard approach were obtained based on the *CS(T)*-data. The corresponding values found from the *A(T)*-data are $T_D$=274(4) K and $T_D$=278(17) K, respectively. The $T_D$-values found for the σ-phase sample using the set of the *CS(T)*-data are 437(7) K and 450(12) K, for the standard and the simple methods, respectively. In turn, the analysis of the *A(T)*-data gave $T_D$=379(5) K and $T_D$=367(30) K, respectively.

The corresponding $T_D$-figures are, within the error limit, in agreement with each other giving thereby evidence that the new method gives correct values of the center shift as well as of the spectral area. The method outlined in this paper is universal i.e. it can be applied to analyze any spectrum, it is much faster than the standard one (practically instant), as it does not involve iteration procedure, and, most of all, it does not require any prior knowledge either on the number of sub spectra, or on their type (singlet, doublet, sextet) or on their relative contributions.

**Acknowledgements**

This work was financed by the Faculty of Physics and Applied Computer Science AGH UST and ACMIN AGH UST statutory tasks within the subsidy from the Ministry of Science and Higher Education, Warszawa.